\documentclass[aps, preprint, showpacs, showkeys,nofootinbib,floatfix]{revtex4-1}

\usepackage{amssymb}
\usepackage{amsmath}
\usepackage{graphicx}

\begin{document}

\title{Testing the Consistency of Gamma Ray Burst Data-set and Supernovae Union2}

\author{Lixin Xu$^{a,b}$\thanks{Corresponding author}}
\email{lxxu@dlut.edu.cn}
\author{Yuting Wang$^a$}
\email{wangyuting0719@163.com}

\affiliation{$^a$Institute of Theoretical Physics, School of Physics
\& Optoelectronic Technology, Dalian University of Technology,
Linggong Road 2$^\sharp$, Dalian, 116024, P. R. China}

\affiliation{$^b$College of Advanced Science \& Technology, Dalian
University of Technology, Linggong Road 2$^\sharp$, Dalian, 116024,
P. R. China}

\begin{abstract}
In this paper, we test the consistency of Gamma Ray Bursts (GRBs)
Data-set and Supernovae Union2 (SNU2) via the so-called {\it
multi-dimensional consistency test} under the assumption that
$\Lambda$CDM model is a potentially correct cosmological model. We
find that the probes are inconsistent with $1.456\sigma$ and
$85.47\%$ in terms of probability. With this observation, it is
concluded that GRBs can be combined with SNU2 to constrain
cosmological models.
\end{abstract}



\maketitle

\section{Introduction}

The cosmic observations of type Ia supernovae (SN Ia) imply that our
universe is undergoing an accelerated expansion
\cite{ref:Riess98,ref:Perlmuter99}. Furthermore, this implication
was confirmed by the observations from cosmic microwave background
radiation \cite{ref:Spergel03,ref:Spergel06} and large scale
structure \cite{ref:Tegmark1,ref:Tegmark2}. However, understanding
the current accelerated expansion of our universe has become one of
the most important issues of modern cosmology. In general, from the
phenomenological point of view, this late time accelerated expansion
of our universe is due to possible modification of gravity theory at
large scale or an exotic extra energy component, dubbed dark energy,
which has negative pressure.

To reveal the nature of the accelerated expansion or properties of
dark energy, one needs more powerful cosmic probes. In the last
decade, the data points of SN Ia (the current SNU2) have amounted to
the number $557$. However, the redshift range of SN Ia is relatively
limited. Of course, higher redshift probes are useful to describe
the evolution of our universe and to reveal the nature of late time
accelerated expansion of our universe and properties of dark energy.
The redshift of GRBs can extend to $z\sim 8.1$ or higher which makes
it as a complementary cosmic probe to SN Ia. But before using GRBs
to constrain cosmological models, the GRBs correlations, which
relate cosmological models and intrinsic properties of GRBs, should
be calibrated first. In general, the GRBs correlations can be
written in a common form of $y=a+bx$ where $a$ and $b$ are the
calibrated parameters, $x$ and $y$ are related to the intrinsic
properties of GRBs and cosmological models, for the details please
see \cite{ref:Schaefer}. However, if one calibrates the GRBs
correlations via a defined cosmological model, say  $\Lambda$CDM
model with $\Omega_{m0}=0.27$, the resulting distance moduli of GRBs
are not independent of the input cosmology model. As a result, the
obtained distance moduli can not be used to constrain any other
cosmological models. The so-called circular problem will be
committed once the above mentioned results are used to constrain any
other cosmological model. Based on this point, the distance modulus
derived by Schaefer \cite{ref:Schaefer} can not be used to constrain
any other cosmological models. So, new methods would be introduced
to overcome this problem. Li, {\it et. al} \cite{ref:Li} put the
GRBs correlation and its cosmological model constraint together as a
whole to fix the calibration parameters and to obtain the best fit
values of the cosmological parameters in different cosmological
models via Markov Chain Monte Carlo (MCMC) method. The lack of the
GRBs calibration makes the GRBs weak to constrain cosmological
models. In fact, the test of the correlations are needed to
guarantee the consistency. In an alternative way, cosmography method
was considered in \cite{ref:cosmography} by parameterizing the
luminosity distance $d_L$ in terms of deceleration parameter $q_0$,
jerk $j_0$ and snap $s_0$ parameters. Clearly, the so-called
circular problem is removed.  However, this Taylor series method is
limited when it is combined with higher redshift data point to
constrain cosmological models. Liang {\it et. al.} \cite{ref:Liang}
used the low redshift SN Ia to calibrate the GRBs correlations and
assumed the correlations were respected at high redshifts. Recently,
this method was reconsidered by Wei \cite{ref:Wei,ref:Wei109}. But,
there would be some problems when GRBs are combined with other
external data sets to constrain cosmological models. Wang presented
a model-independent distance measurement $\bar{r}_p(z_i)$ (Eq.
(\ref{eq:rp}) of this paper) from GRBs calibrated internally
\cite{ref:wang}, where $z_i$ are the redshifts of GRBs. The main
points of Wang's method are that the resulted distance measurement
$\bar{r}_p(z_i)$ is cosmological model in-dependent. The values of
correlation parameters $a$ and $b$ are not used directly but the
statistical errors of correlation parameters $\sigma_a$, $\sigma_b$
and systematic error $\sigma_{sys}$ are. This is because the
$1\sigma$ error bars of $a$, $b$ and systematic error are almost the
same in $\Lambda$CDM with different values of $\Omega_{m0}$, though
the values of $a$ and $b$ are different. Then, the cosmic constraint
from GRBs is obtained in terms of a set of model-independent
distance measurements. The merits of this method are listed as
follows: (i) the constraint from GRBs is in a cosmological model
independent way. So, it can be used to constrain other cosmological
models. (ii) It is not calibrated by any other external data sets.
It will not suffer any consistent problem when it is combined with
other data sets as cosmic constraints. (iii) The cosmological model
independent calibration is done first. (iv) Though the absolute
calibration of GRBs is not known, the slopes of GRBs correlations
can be used as cosmological constraints. Clearly, the drawback is
clear that the constrained result is not tighter than the one
calibrated by using SN Ia. But, if we have enough data points of
GRBs, this problem will be overcome. Because the slopes of GRBs
correlations are considered alone, this may make the GRBs not very
powerful.

Following the method proposed by Wang \cite{ref:wang}, Xu obtained
$N=5$ model-independent distances data sets and their covariance
matrix by using $109$ GRBs via Amati's $E_{p,i}-E_{iso}$ correlation
\cite{ref:GRBsXu}. These five model-independent distances data
points have been used to constrain cosmological model \cite{DGP:XU}.
However, the consistency of the obtained five data sets via Amati's
correlation with other cosmic probes must be checked to guarantee
the reliability of GRBs. With this motivation, we will test the
consistency or inconsistency of SNU2 with these five data sets
derived from GRBs via the so-called {\it multi-dimensional
consistency test} which will be reviewed briefly in section
\ref{sec:method}, for the details please see \cite{ref:mulititest}.

This paper is structured as follows. In section \ref{sec:DM}, the
SNU2, the five data sets derived from GRBs and the method to
constrain dark energy model are presented. Also, the
multi-dimensional consistency test is reviewed briefly. Section
\ref{sec:conclusion} is the concluding remark.

\section{Data-sets and Method}\label{sec:DM}

\subsection{Type Ia Supernovae}

Recently, SCP (Supernova Cosmology Project) collaboration released
their Union2 dataset which consists of 557 SN Ia \cite{ref:SN557}.
The distance modulus $\mu(z)$ is defined as
\begin{equation}
\mu_{th}(z)=5\log_{10}[\bar{d}_{L}(z)]+\mu_{0},
\end{equation}
where $\bar{d}_L(z)$ is the Hubble-free luminosity distance $H_0
d_L(z)/c=H_0 d_A(z)(1+z)^2/c$, with $H_0$ the Hubble constant,
and $\mu_0\equiv42.38-5\log_{10}h$ through the re-normalized quantity $h$ as $H_0=100 h~{\rm km
~s}^{-1} {\rm Mpc}^{-1}$. Where
$d_L(z)$ is defined as
\begin{equation}
d_L(z)=(1+z)r(z),\quad r(z)=\frac{c}{H_0\sqrt{|\Omega_{k}|}}{\rm
sinn}\left[\sqrt{|\Omega_{k}|}\int^z_0\frac{dz'}{E(z')}\right]
\end{equation}
where $E^2(z)=H^2(z)/H^2_0$. Additionally, the observed distance
moduli $\mu_{obs}(z_i)$ of SN Ia at $z_i$ are
\begin{equation}
\mu_{obs}(z_i) = m_{obs}(z_i)-M,
\end{equation}
where $M$ is their absolute magnitudes.

For the SN Ia dataset, the best fit values of the parameters $p_s$
can be determined by a likelihood analysis, based on the calculation
of
\begin{eqnarray}
\chi^2(p_s,M^{\prime})&\equiv& \sum_{SN}\frac{\left\{
\mu_{obs}(z_i)-\mu_{th}(p_s,z_i)\right\}^2} {\sigma_i^2}  \nonumber\\
&=&\sum_{SN}\frac{\left\{ 5 \log_{10}[\bar{d}_L(p_s,z_i)] -
m_{obs}(z_i) + M^{\prime} \right\}^2} {\sigma_i^2}, \label{eq:chi2}
\end{eqnarray}
where $p_s=\{\Omega_{m0}\}$ denotes the model parameter and
$M^{\prime}\equiv\mu_0+M$ is a nuisance parameter which includes the
absolute magnitude and the parameter $h$. The nuisance
  parameter $M^{\prime}$ can be marginalized over
analytically \cite{ref:SNchi2} as
\begin{equation}
\bar{\chi}^2(p_s) = -2 \ln \int_{-\infty}^{+\infty}\exp \left[
-\frac{1}{2} \chi^2(p_s,M^{\prime}) \right] dM^{\prime},\nonumber
\label{eq:chi2marg}
\end{equation}
resulting to
\begin{equation}
\bar{\chi}^2 =  A - \frac{B^2}{C} + \ln \left( \frac{C}{2\pi}\right)
, \label{eq:chi2mar}
\end{equation}
with
\begin{eqnarray}
A&=&\sum_{i,j}^{SN}\left\{5\log_{10}
[\bar{d}_L(p_s,z_i)]-m_{obs}(z_i)\right\}\cdot {\rm Cov}^{-1}_{ij}\cdot
\left\{5\log_{10}
[\bar{d}_L(p_s,z_j)]-m_{obs}(z_j)\right\},\nonumber\\
B&=&\sum_i^{SN} {\rm Cov}^{-1}_{ij}\cdot \left\{5\log_{10}
[\bar{d}_L(p_s,z_j)]-m_{obs}(z_j)\right\},\nonumber \\
C&=&\sum_i^{SN} {\rm Cov}^{-1}_{ii},\label{eq:SNsyserror}
\end{eqnarray}
where ${\rm Cov}^{-1}_{ij}$ is the inverse of covariance matrix with or without systematic
errors. One can find the details in
Ref. \cite{ref:SN557} and the web site
\footnote{http://supernova.lbl.gov/Union/} where
the covariance matrix with or without systematic errors are included. Relation (\ref{eq:chi2}) has a minimum at the nuisance parameter
value $M^{\prime}=B/C$, which contains information of the values of
$h$ and $M$. Therefore, one can extract the values of $h$ and $M$
provided the knowledge of one of them. Finally, the
expression
\begin{equation}
\chi^2_{SN}(p_s,B/C)=A-(B^2/C),\label{eq:chi2SN}
\end{equation}
which coincides to Eq. (\ref{eq:chi2mar}) apart from a constant, is
often used in the likelihood analysis
\cite{ref:smallomega,ref:SNchi2}. Thus in this case the results will
not be affected by a flat $M^{\prime}$ distribution. It is worth
noting that the results will be different with or without the
systematic errors. In this work, all results are obtained with
systematic errors.

\subsection{Gamma Ray Bursts}

Following \cite{ref:Schaefer}, we consider the well-known Amati's
$E_{p,i}-E_{iso}$ correlation \cite{ref:Amati'srelation,r16,r17,r18}
in GRBs, where $E_{p,i}=E_{p,obs}(1+z)$ is the cosmological
rest-frame spectral peak energy, and $E_{iso}$ is the isotropic
energy
\begin{equation}
E_{iso}=4\pi d^2_LS_{bolo}/(1+z)
\end{equation}
in which $d_L$ and $S_{bolo}$ are the luminosity distance and the
bolometric fluence of the GRBs respectively. Following
\cite{ref:Schaefer}, we rewrite the Amati's relation as
\begin{equation}
\log\frac{E_{iso}}{{\rm erg}}=a+b\log\frac{E_{p,i}}{300{\rm
keV}}.\label{eq:calib}
\end{equation}

In \cite{ref:wang}, Wang defined a set of model-independent distance
measurements $\{\bar{r}_p(z_i)\}$:
\begin{equation}
\bar{r}_p(z_i)\equiv\frac{r_p(z)}{r_p(z_{0})},\quad r_p(z)\equiv
\frac{(1+z)^{1/2}}{z}\frac{H_0}{c}r(z),\label{eq:rp}
\end{equation}
where $r(z)=d_L(z)/(1+z)$ is the comoving distance at redshift $z$, and
$z_{0}$ is the lowest GRBs redshift. Then, the cosmological model
can be constrained by GRBs via
\begin{eqnarray}
\chi^2_{GRBs}(p_s)&=&[\Delta\bar{r}_p(z_i)]\cdot(Cov^{-1}_{GRB})_{ij}\cdot[\Delta\bar{r}_p(z_i)],\label{eq:chi2GRB}\\
\Delta\bar{r}_p(z_i)&=&\bar{r}^{data}_p(z_i)-\bar{r}_p(z_i),
\end{eqnarray}
where $\bar{r}_p(z_i)$ is defined by Eq. (\ref{eq:rp}) and
$(Cov^{-1}_{GRB})_{ij},i,j=1...N$ is the covariance matrix. In this
way, the constraints from observational GRBs data are
projected into the relative few quantities $\bar{r}_p(z_i),i=1...N$.

Following the method proposed by Wang \cite{ref:wang}, Xu obtained
$N=5$ model-independent distances data sets and their covariance
matrix by dividing $109$ GRBs into five bins via Amati's
$E_{p,i}-E_{iso}$ correlation \cite{ref:GRBsXu}. The resulting
model-independent distances and covariance matrix from $109$ GRBs
are shown below in Tab. \ref{tab:distance}
\begin{table}[htbp]
\begin{center}
\begin{tabular}{c|c|c|c|c}
\hline\hline
 & $z$ & $\bar{r}^{data}_p(z)$ & $\sigma(\bar{r}_p(z))^+$ &  $\sigma(\bar{r}_p(z))^-$\\ \hline
$0$ & $\quad0.0331\quad$ & $\quad1.0000\quad$ & $-$ & $-$  \\
$1$ & $1.0000$  & $0.9320$ & $0.1711$ & $0.1720$ \\
$2$ & $2.0700$  & $0.9180$ & $0.1720$ & $0.1718$ \\
$3$ & $3.0000$  & $0.7795$ & $0.1630$ & $0.1629$ \\
$4$ & $4.0480$  & $0.7652$ & $0.1936$ & $0.1939$ \\
$5$ & $8.1000$  & $1.1475$ & $0.4297$ & $0.4389$ \\
  \hline\hline
\end{tabular}
\end{center}
\caption{\label{Tab:datapoints} Distances measured from $109$ GRBs
via Amati's correlation with $1\sigma$ upper and lower uncertainties
\cite{ref:GRBsXu}. $z_{0}=0.0331$ as lowest redshift was
adopted.}\label{tab:distance}
\end{table}
and Eq. (\ref{eq:covM}). The $\{\bar{r}_p(z_i)\}(i=1,...,5)$
correlation matrix is given by
\begin{eqnarray}
&&(\overline{Cov}_{GRB})= \left(
\begin{array}{ccccc}
$1.0000$ & $0.7780$ & $0.8095$ & $0.6777$ & $0.4661$ \\
$0.7780$ & $1.0000$ & $0.7260$ & $0.6712$ & $0.3880$ \\
$0.8095$ & $0.7260$ & $1.0000$ & $0.6046$ & $0.5032$ \\
$0.6777$ & $0.6712$ & $0.6046$ & $1.0000$ & $0.1557$ \\
$0.4661$ & $0.3880$ & $0.5032$ & $0.1557$ & $1.0000$
\end{array}
\right),
\end{eqnarray}
and the covariance matrix is given by
\begin{equation}
(Cov_{GRB})_{ij}=\sigma(\bar{r}_p(z_i))\sigma(\bar{r}_p(z_j))(\overline{Cov}_{GRB})_{ij},\label{eq:covM}
\end{equation}
where
\begin{eqnarray}
\sigma(\bar{r}_p(z_i))=\sigma(\bar{r}_p(z_i))^+, \quad {\rm if}\quad
\bar{r}_p(z)\geq
\bar{r}_p(z)^{data}; \\
\sigma(\bar{r}_p(z_i))=\sigma(\bar{r}_p(z_i))^-, \quad {\rm if}\quad
\bar{r}_p(z)< \bar{r}_p(z)^{data},
\end{eqnarray}
the $\sigma(\bar{r}_p(z_i))^+$ and $\sigma(\bar{r}_p(z_i))^-$ are
the $1\sigma$ errors listed in Tab. \ref{Tab:datapoints}.

\subsection{Method: Multi-dimensional Consistency Test}\label{sec:method}

\begin{figure}[!htbp]
\includegraphics[width=8cm]{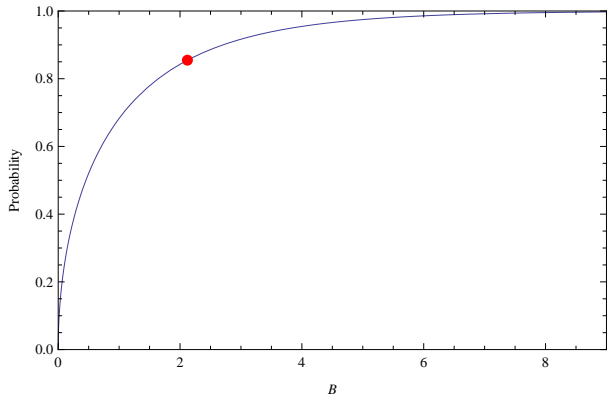}
\caption{The relation between $B$ and probability where $\nu=1$ is
adopted, where the red point denotes the probability $85.47\%$ at
$B=2.121$.} \label{fig:PB}
\end{figure}

In Ref. \cite{ref:mulititest}, the multi-dimensional consistency
test of probes was considered. If we consider $M$ parameters and $N$
probes, the method to test the consistency is to minimize the
$\chi^2(\lambda_\alpha)$ with respect to $\lambda_\alpha$,
\begin{equation}
\chi^2(\lambda_\alpha)=\sum^M_{i=1}\sum^N_{\alpha,\beta=1}(\lambda_{\alpha}-\lambda^{(i)}_{\alpha})[C^{(i)}]^{-1}_{\alpha\beta}(\lambda_{\beta}-\lambda^{(i)}_{\beta})\label{eq:chi2lambda},
\end{equation}
where $\lambda^{(i)}_{\alpha}$ is the best fit value returned from
the $i$th probe with covariance matrix $C^{(i)}_{\alpha\beta}$, and
$\lambda_{\alpha}$ is a random point in cosmological space. In this
case, the value of $\lambda_\alpha$ at the minimum $\chi^2$ is the
best fit value. The goodness of fit is quantified by the value of
$\chi^2$ in the standard way, i.e., by checking the expectation
value $<\chi^2_{min}>=\nu$ where $\nu$ is the degrees of freedom.
For example, in our case, we consider the $\Lambda$CDM model with
$\Omega_{m0}$ as a free model parameter ($M=1$) and two probes
(N=2): SN and GRBs. So, the degrees of freedom ($\nu=N-M$) are
$2-1=1$. The expectation value of $\chi^2_{min}$ would be $1$.
However, the value $<\chi^2_{min}>=\nu+B$ will be returned, where
$B>0$ denotes the possible deviation from $\nu$. Then, the
consistency can be concluded by the value of $d_{\sigma}$ via the
formula
\begin{equation}
1-\Gamma(\nu/2,B/2)/\Gamma(\nu/2)={\rm
Erf}(d_{\sigma}/\sqrt{2}).\label{eq:distance}
\end{equation}
The larger value of $d_{\sigma}$ denotes better inconsistency
between the probes. For example, a difference $B=9$ tells us the two
probes are inconsistent with $99.7\%$ ($3\sigma$) in $\Lambda$CDM
model. For convenience, we show the relation between $B$ and
probability in Fig. \ref{fig:PB} where $\nu=1$ is adopted.

As described above, we firstly find the corresponding minimum
$\chi^2_{min}$ values with SN and GRBs via Markov Chain Monte Carlo
(MCMC) method. Our code is based on the publicly available {\bf
CosmoMC} package \cite{ref:MCMC}. The results are shown in Tab.
\ref{tab:results}. Via the formula (\ref{eq:chi2lambda}), we find
the minimum value of $\chi^2(\Omega_{m0})$ is $3.121$ with the best
fit value of $\Omega_{m0}=0.287$, where the covariance matrix
$C^{(i)}_{\alpha\beta}$ is given
\begin{equation}
C^{(i)}_{\alpha\beta}=\sigma^{i}_{\alpha}(\lambda^i)\sigma_{\beta}(\lambda^i)\delta_{\alpha\beta},\label{eq:covlambda}
\end{equation}
here
\begin{eqnarray}
\sigma^{i}_{\alpha}(\lambda^i)=\sigma^{i}_{\alpha}(\lambda^i)^+, \quad {\rm if}\quad
\lambda_{\alpha}\geq
\lambda^{i}_{\alpha}; \\
\sigma^{i}_{\alpha}(\lambda^i)=\sigma^{i}_{\alpha}(\lambda^i)^-, \quad {\rm if}\quad
\lambda_{\alpha}<
\lambda^{i}_{\alpha},
\end{eqnarray}
and the correlation between SN and GRBs is zero. However, the
expected value of $\chi^2(\Omega_{m0})$ should be $\nu=1$. Then the
returned deviation $B$ is $2.121$. From Eq. (\ref{eq:distance}), one
finds that the probes are inconsistent with $d_{\sigma}=1.456\sigma$
and $85.47\%$ in terms of probability.

\begin{table}
\begin{center}
\begin{tabular}{cc|    cc    cc|  cc }
\hline\hline
Datasets & & Parameters & & $\chi^2_{min}/{\rm d.o.f}$ && $\Omega_{m0}$
\\ \hline
SNU2 && 1 && 530.722/556 && $0.274^{+0.0386}_{-0.0358}$ \\
GRB  && 1 &&  3.0354/4 && $0.620^{+0.313}_{-0.192}$  \\
\hline\hline
\end{tabular}
\caption{The results of $\chi^2_{min}$, $\Omega_{m0}$ with
$1\sigma$ regions are listed, where SN systematic errors are included. ${\rm d.o.f}$ denotes the degrees of
freedom.}\label{tab:results}
\end{center}
\end{table}

\section{Conclusion}\label{sec:conclusion}

The consistency of SN Ia and GRBs is checked by the so-called {\it
multi-dimensional consistency test} under the assumption that
$\Lambda$CDM model is a correct cosmological model. We find that the
inconsistency of SNU2 and GRBs is about $1.456\sigma$ and $85.47\%$
in terms of probability. So, we can conclude that the five GRBs data
points are consistent with SNU2 at the level above $1.456\sigma$.
These five GRBs data sets can be combined with SNU2 to constrain
other cosmological models.

\acknowledgements{This work is supported by NSF (10703001), SRFDP
(20070141034) of P.R. China and the Fundamental Research Funds for
the Central Universities (DUT10LK31). We appreciate the anonymous
referee's invaluable help to improve this work.}

\end{document}